\documentstyle[11pt,emulateapj5]{aastex}

\def\kmsMpc{\mbox{\rm \,km\,s}^{-1}{\rm Mpc}^{-1}}

\begin{document}

\title{Intrinsic Degeneracy of Gravitational Lens Time Delays: 
the case for a simple quadruple system in $\Lambda$CDM cosmology}
\author{HongSheng Zhao and Bo Qin\altaffilmark{1,2}}
\submitted{To be published in ApJ, 582, 000 (2003)}
\altaffiltext{1}{National Astronomical Observatories, Chinese Academy of Science, Beijing 100012, PRC}
\altaffiltext{2}{Institute of Astronomy, Madingley Road, Cambridge, CB3 0HA, U.K. (zhao,qinbo@ast.cam.ac.uk)}
\begin{abstract}
A degeneracy in strong lens model is shown analytically.  
The observed time delays and quasar image positions might {\it not} uniquely 
determine the concentration and 
the extent of the lens galaxy halo mass distribution.
Simply hardwiring the Hubble constant ($H_0$) 
and the cosmology ($\Omega, \Lambda$) 
to the standard $\Lambda$CDM cosmology values
might {\it not} fully lift this degeneracy, which exists rigourously
even with very accurate data.  Equally good fits to the images could be found
in lens mass models with either a mostly Keplerian or a flat rotation curve.  
This degeneracy in mass models makes the task of getting reliable
$H_0$ and $\Lambda$ from strong lenses even more daunting.
\end{abstract}

\keywords{cosmological parameters---dark matter---distance scale
---gravitational lensing}

\section{Introduction}

One of the promises of gravitational lenses is to measure the Hubble
constant $H_0$ (Refsdal 1964) from the observed time delays among
the images of a variable background quasar source lensed by a
foreground galaxy.  Given a model for the spatial distributions of 
the stars and dark matter in the
lens galaxy, the time delay $\Delta t_{\rm obs}$ multiplied by the speed of
light $c$ is simply proportional to the absolute distances to the lens
and the source, hence $c\Delta t_{\rm obs}$ scales with the size
of the universe $c/H_0$.  While the
time delays are now routinely measured for many systems (see Schechter
2000), a reliable determination of $H_0$ has been hampered to some
extent by the intrinsic degeneracy in models of the dark matter
potential of the lens (Williams \& Saha 2000; Saha 2000; Zhao \& Pronk
2001).  The general trend is that a model with a dense dark matter halo
gives a small $H_0$ with 
\begin{equation}\label{trend}
H_0 \Delta t_{\rm obs} \propto 
\left[2\Sigma_{\rm crit}(z_l,z_s)-\Sigma_{\rm *}(R_E)-\Sigma_{\rm h}(R_E)\right],
\end{equation}
where $\Sigma_{\rm crit}(z_l,z_s)$
is the critical density for the lens at redshift $z_l$ and the source at $z_s$,
and $\Sigma_{\rm *}(R_E)$ or $\Sigma_{\rm h}(R_E)$ 
is the typical surface density of luminous or dark matter at 
within the Einstein ring $R_E$
(Falco, Gorenstein, \& Shapiro 1985; Kochanek 2002).  

Given that the value of $H_0$ is now
well constrained by other independent methods, such as the HST Key
Project (Freedman et al. 2001), it is interesting to
reverse the angle of the question, and use
equation$~$(\ref{trend}) to put more stringent
constraint on the dark matter potential of the lens.  
More specifically in this {\sl Letter}, we would like to ask the question: 
how narrow is the allowed parameter space for the lens dark halo
which is consistent with a given set of images, time delays, cosmology
and $H_0$? In the interest of clarity, 
we will consider only analytical lens models 
with a simplified geometry for a hypothetical image and lens system.
We believe our arguments should 
apply qualitatively to real galaxy lenses as well, and a more detailed
application to the quadruple system PG1115+080 is given 
in a follow-up paper (Zhao \& Qin 2002).

\section{Analytical Lens Models for Stars and Halo}

For simplicity 
we assume the four images form a perfectly symmetric Einstein cross with
the time delay minima on the y-axis at  $\pm R_E$ (radian) from the lens
center, and the saddle point images on the x-axis at $\pm qR_E$ (radian).
For generality we rescale the images, taking
the radius $R_E$ as unity in a new $XY$ coordinate system, so
the four images are given by
\begin{equation}
[X,Y]_{\rm minima} = [0,\pm 1],~ 
[X,Y]_{\rm saddle} = [\pm q,0].
\end{equation}
This image configuration implies that 
a source $(X_s,Y_s)=(0,0)$, exactly behind a spherical lens
plus a linear external shear symmetric to the X and Y axes.  

All lensing properties can be derived from the time delay surfaces.
For our model with spherical stellar lens $\phi_*(R)$ plus halo $\phi_h(R)$
and a linear shear of amplitude $\gamma_1$, 
the time delay contours are determined by 
a dimensionless time delay $\tau(X,Y)$ given by
\begin{eqnarray}
\tau(X,Y) & = & t \cdot \omega(H_0,\Omega,z_l,z_s) R_E^{-2}  \\
&=& {X^2 +Y^2\over 2}- {\gamma_1 (X^2-Y^2)\over 2}- \phi_*(R) - \phi_h(R),
\end{eqnarray}
where $R=\sqrt{X^2+Y^2}$ is the cylindrical radius,
and $\omega(H_0,\Omega,z_l,z_s)$ is a constant containing 
all the dependence on the cosmological density parameter $\Omega$ and
the lens and source redshifts $z_l,z_s$; crudely speaking 
$\omega \sim H_0$ for typical
lens and source redshifts in the $\Lambda$CDM cosmology.  

To illustrate the non-uniqueness, we need to 
find reasonable lens models with similar time delay surface.
Let's consider the following example of lensing potential for the stars,
\begin{equation}
\phi_*(R) = {m_0 \over \alpha}\ln \left(1+{R^\alpha \over a^\alpha}\right) 
\end{equation}
where $m_0$ is the total stellar mass enclosed, $a$ is the half mass radius,
and $2-\alpha$ specifies the cuspiness of the stellar distribution.
For the halo we use a nearly isothermal potential
\begin{equation}
\phi_h(R) = b_0 R - {\delta \over 2} 
\left[(1,R^2)_{\rm min} +\ln (1,R^2)_{\rm max}+\ln^2 (1,R^2)_{\rm max} \right]
\end{equation}
where $\sqrt{b_0}$ 
is proportional to the terminal velocity of the halo rotation curve.

Note that the function $\phi_h$ is smoothly connected at the two sides of 
the radius $R=1$, which is chosen such that it is just outside the images.
The parameter $\delta$ is a dimensionless tunable parameter to adjust the halo
contribution to the surface density at $R=1$.
The surface density can be computed as
\begin{equation}
\kappa(R) = 1-{1 \over 2} \nabla^2 \tau,
\end{equation}
so the densities for the stars and the halo are given by
\begin{equation}
\kappa_*(R) =  {\alpha m_0 a^\alpha\over 2 R^{2-\alpha} (R^\alpha+a^\alpha)^2},
\qquad \kappa_h(R) =  {b_0 \over 2 R}- {\delta \over [1,R^2]_{\rm max}}.
\end{equation}
Note that the stars have an inner cusp $2-\alpha$,
and the halo density $\kappa_h$ 
is continuous across the break $R=1$, and positive 
everywhere if $b_0/\delta >2$.   

By increasing $\delta$ we can lower the mean densities at the images,
hence creating the effect of a negative mass sheet.  This will not
affect the positions of the images, but can increase the time delay 
$\Delta \tau_{\rm obs}
\propto H_0\Delta t_{\rm obs}$ between the images.  As we will see, this can
result in a larger $H_0$ to be consistent with the currently favored 
high value of $H_0 \sim 70 \kmsMpc$.

The deflection strength 
\begin{equation}
b(R) \equiv {d \phi \over d R} \equiv {M_*(<R)+M_h(<R) \over R}
\end{equation}
is effectively the rotation curve squared.  A flat rotation curve corresponds
to a constant deflection strength.  For our model, the stellar and halo masses
enclosed inside radius $R$ are given by
\begin{equation}
M_* = {m_0R^\alpha \over R^\alpha+a^\alpha},~~
M_h =b_0R - \delta \left[(1,R^2)_{\rm min}+\ln (1,R^2)_{\rm max}\right].
\end{equation}
As we can see the stars converge to a finite mass $m_0$ at infinity and 
the halo dominates stars and approaches to a finite deflection (terminal velocity $\propto \sqrt{b_0}$).

\section{Results}

Let's consider a typical lensing system 
in a standard $\Lambda$CDM cosmology with 
\begin{equation}
H_0=70 \kmsMpc,\qquad \Omega=1-\Lambda=0.3
\end{equation}
with the lens and source redshifts
\begin{equation}
z_l=0.5,\qquad z_s=2,
\end{equation}
and a time delay $\Delta t_{\rm obs}$ 
\begin{equation}
\left(t_{\rm saddle}^{\rm obs}-t_{\rm mimima}^{\rm obs}\right)
=28 \, {\rm day}\left[\left({R_E\over1\arcsec}\right)^2-
\left({qR_E\over1\arcsec}\right)^2\right],~q =0.5\,
\end{equation}
between the saddle point image at $(qR_E,0)$ and the minima image at $(0,R_E)$.

The above cosmology specifies the distances to the lens and the source, hence 
the time delay normalization constant 
\begin{equation}
\omega(H_0=70,\Omega=1-\Lambda=0.3,z_l=0.5,z_s=2) \sim 100 \kmsMpc.
\end{equation}

The observed time delay then set the following 
constraint on the lens model,
\begin{eqnarray}\label{fitdelay}
\Delta \tau_{\rm obs} & = & \tau(0.5,0)-\tau(0,1) \\
& = & {\left(t_{\rm saddle}^{\rm obs}-t_{\rm mimima}^{\rm obs}\right)
\over R_E^2 }  \omega \\
&\sim & 28(1-q^2){\rm \, day}/\sq\arcsec \times 
100\kmsMpc \sim 0.22 \qquad.
\end{eqnarray}

We solve for the parameters of the lens and the shear to reproduce
the four images at $(\pm q,0)$, $(0,\pm 1)$ exactly.  
The images form at the extreme points of the time delay surface, hence
we have the  additional constraints
\begin{equation}\label{fitimage}
{\partial \tau \over \partial X} = 0,~
{\partial \tau \over \partial Y} = 0,~\mbox{\rm at $(X,Y)=(0.5,0)$ and $(0,1)$.}
\end{equation}

\begin{figure}
\plottwo{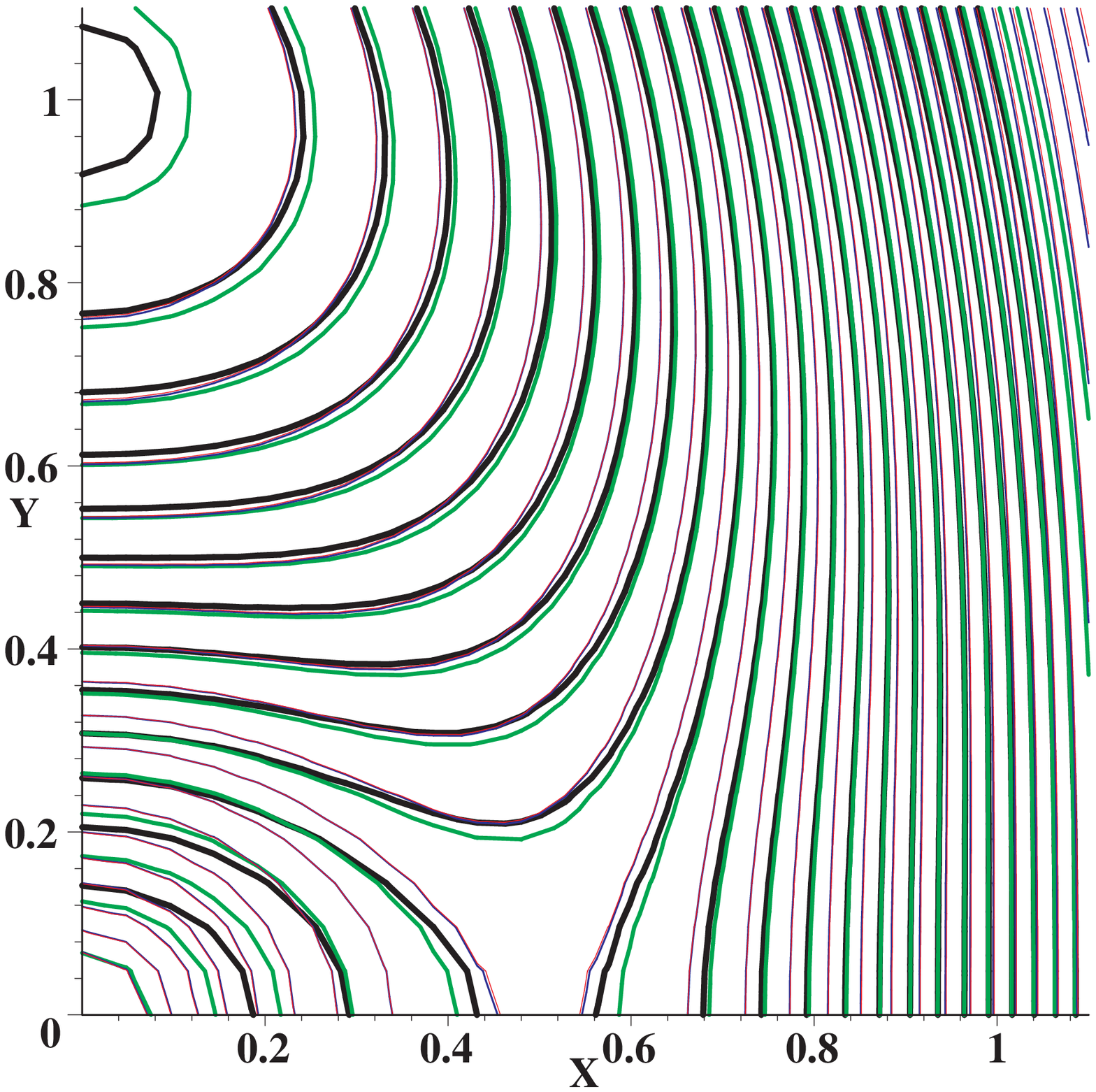}{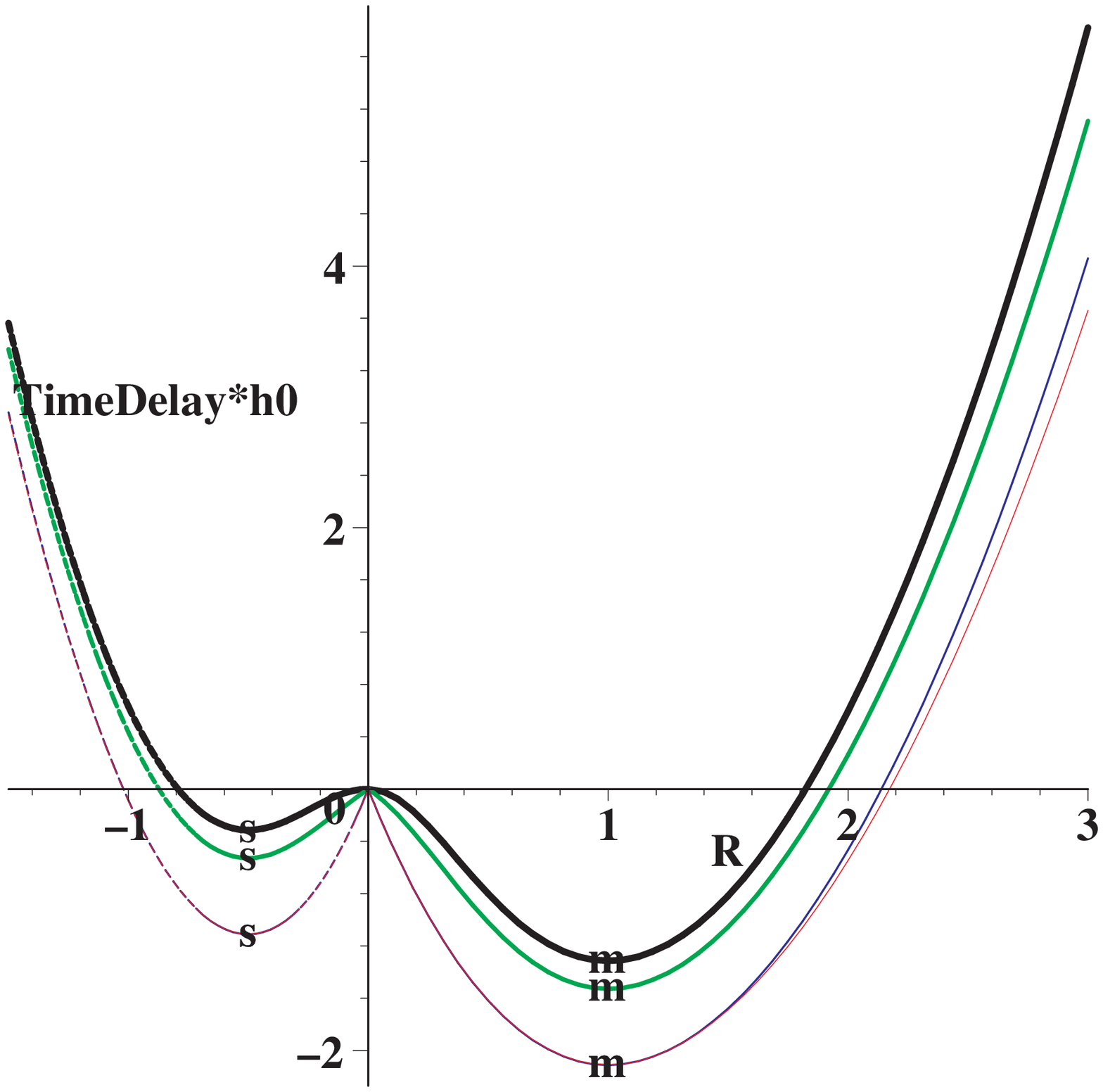}
\caption{
Panel (a): Time delay contours in the XY quadrant 
in intervals of 3 days for the four lens models.
All models reproduce the same minima and saddle image positions and time delay.
Panel (b): Cuts of the time delay surfaces ${t(R)\over 
\left(t_{\rm saddle}^{\rm obs}-t_{\rm mimima}^{\rm obs}\right)h_0^{-1}}$
as a function of the radius $R$
along the radial direction from the lens to the time delay minima
(marked by ``m'' on solid curves to the right) 
and from the lens to the saddle image (marked by ``s'' on dashed curves
to the negative $R$ side) for all four lens models.
All models use $h_0=0.7$ and 
the same time delay between the images ``s'' and ``m''. 
}
\label{ctt}
\end{figure}

It turns out that the allowed models follow a three-parameter
sequence, say $(\delta,a,\alpha)$.  The parameters of a few
representative models are listed in Table 1.  The images appear
rigourously at the same locations for all four models, 
which are the minima and saddle points of
the arrival time surfaces (cf. Figure~\ref{ctt}a), 
and the relative time delay is identical as well 
for all four lens models (cf. Figure~\ref{ctt}b).
But the mass distributions in the four models are far from similar
(cf. Figure~\ref{vk}a and Figure~\ref{vk}b):  lensIII and lensIV
correspond to systems with a finite stellar core with a very 
typical half-mass 
radius $R_e$ about half of the Einstein radius $R_E$; lensIV is purely
in stars, and has no halo.  Both lensI and lensII correspond to systems
with a strong stellar cusp dominating the halo at small radii; 
lensII has a bigger stellar component with a half-mass radius at one $R_E$.

\subsection{The cause of degeneracy}

The above degeneracy 
is a variation of the well-known mass-sheet degeneracy.
The latter implies that
we can increase $H_0$ by reducing the surface density 
$\kappa={\Sigma \over 2\Sigma_{\rm crit}}$ 
at the Einstein radius $R_E$ (cf. equation$~$(\ref{trend})), e.g., by either 
scaling down the stellar mass $m_0$ or scaling down the isothermal 
dark halo mass $b_0 r$.  But if we reduce
the stellar mass $m_0$ and increase the halo mass $b_0 r$
{\sl simultaneously} such that we keep
$\kappa_*(R_E)+\kappa_h(R_E)$ constant (between 0.10 to 0.13, cf. Table 1), 
then $H_0 \Delta t_{\rm obs}$ will not change, 
only the terminal velocity of the rotation curve $\sqrt{b_0}$ is raised.
The result is a rigorous degeneracy of the stellar vs. 
halo mass distribution, 
insensitive to $H_0$ and the observed time delay $\Delta t_{\rm obs}$.

\subsection{Break the degeneracy from observable flux ratio and lens 
light profile?}

\begin{figure}
\plotone{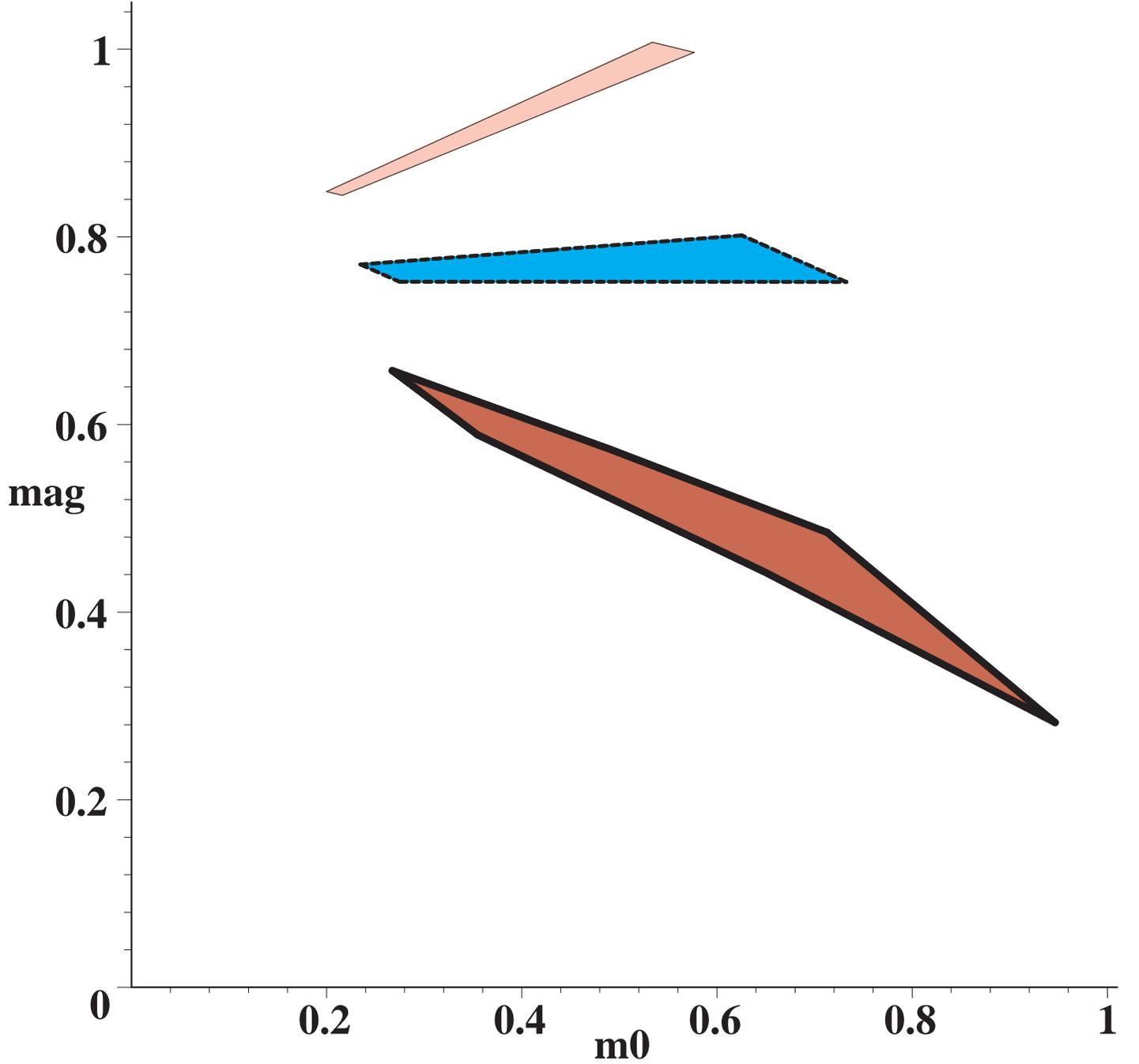}
\caption{The relative magnitude of images
as a function of stellar mass of the model $m_0$.  
From top to down (color coded as pink, cyan and orange)
the painted polygons correspond to stellar cusps of slope $2-\alpha=1,0.5,0$.
Clockwise from the rightmost corner the four corners of 
each polygon correspond to 
$[\delta,a]=[0,0.55],[0.2,0.55],[0.2,0.5],[0,0.5]$.
If these models are extended to $\delta=0.3$,
then all models merger to a point on the vertical axis, 
i.e., a pure-halo model with $m_0 \sim 0$.}
\label{mag}
\end{figure}

The flux ratio of the saddle image and the minima image can 
in princinple differentiate some of the models, as shown by Fig.~\ref{mag}.  
But assuming a reasonable $\pm 0.1$ mag error with the magnitude,
and a 10\% error with the effective radius, 
most of the models are in fact degenerate.  In particular,
it will be difficult to constrain the mass of the stellar component $m_0$.

Observations of a well-resolved stellar lens can fix
the effective radius $a=R_e/R_E$ and perhaps the cuspiness $2-\alpha$.
This reduces the available lens models drasticly.  
Nonetheless, the degeneracy between lensIII and lensIV
implies that it is still problematic to differentiate between models with
dark halo and models without.  To break the degeneracy one needs at least
an accurate measurement of the external shear $\gamma_1$ to 10\% level
(cf. the parameters of lensIII and lensIV in Table 1), perhaps by
a combination of strong lensing and weak lensing data.

\section{Conclusion and Comparison with Earlier Models}

It is possible to construct many very different models with positive,
smooth and monotonic surface densities to fit the image positions.
There are also no extra images.  These models fit the same images,
time delay, $H_0$ and cosmology.  Some fit the same lens light profile
and image flux ratio as well.  Hence the models are virtually {\sl
indistinguishable} from lensing data.  There are severe degeneracies
in inverting the data of a perfect Einstein cross to the lens models,
even if given the Hubble constant and cosmology.  These rigourous
findings with analytical models are also consistent with earlier
numerical models of Saha \& Williams (2001) and semi-analytical models
of Zhao \& Pronk (2001).

\begin{figure}
\plottwo{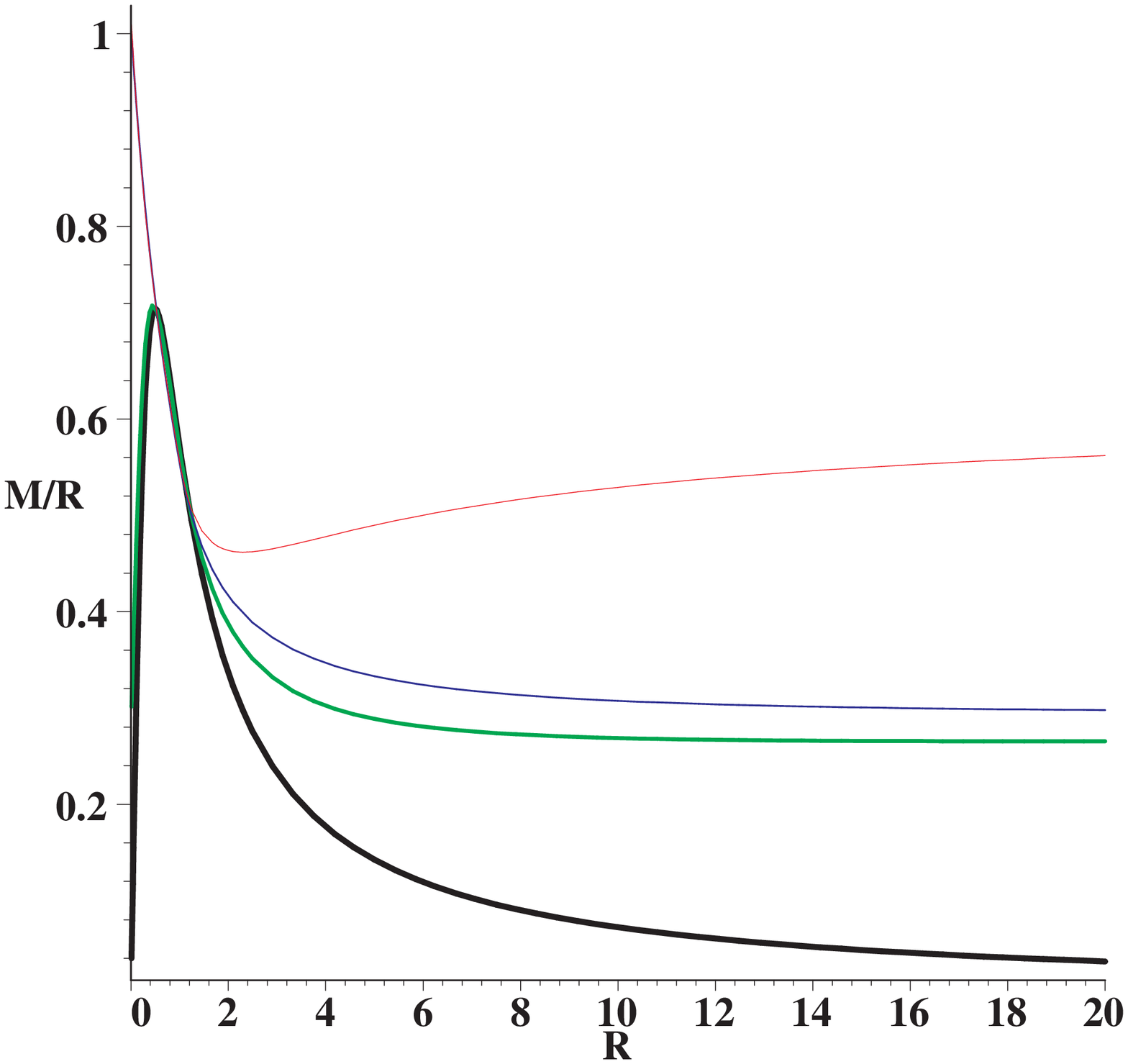}{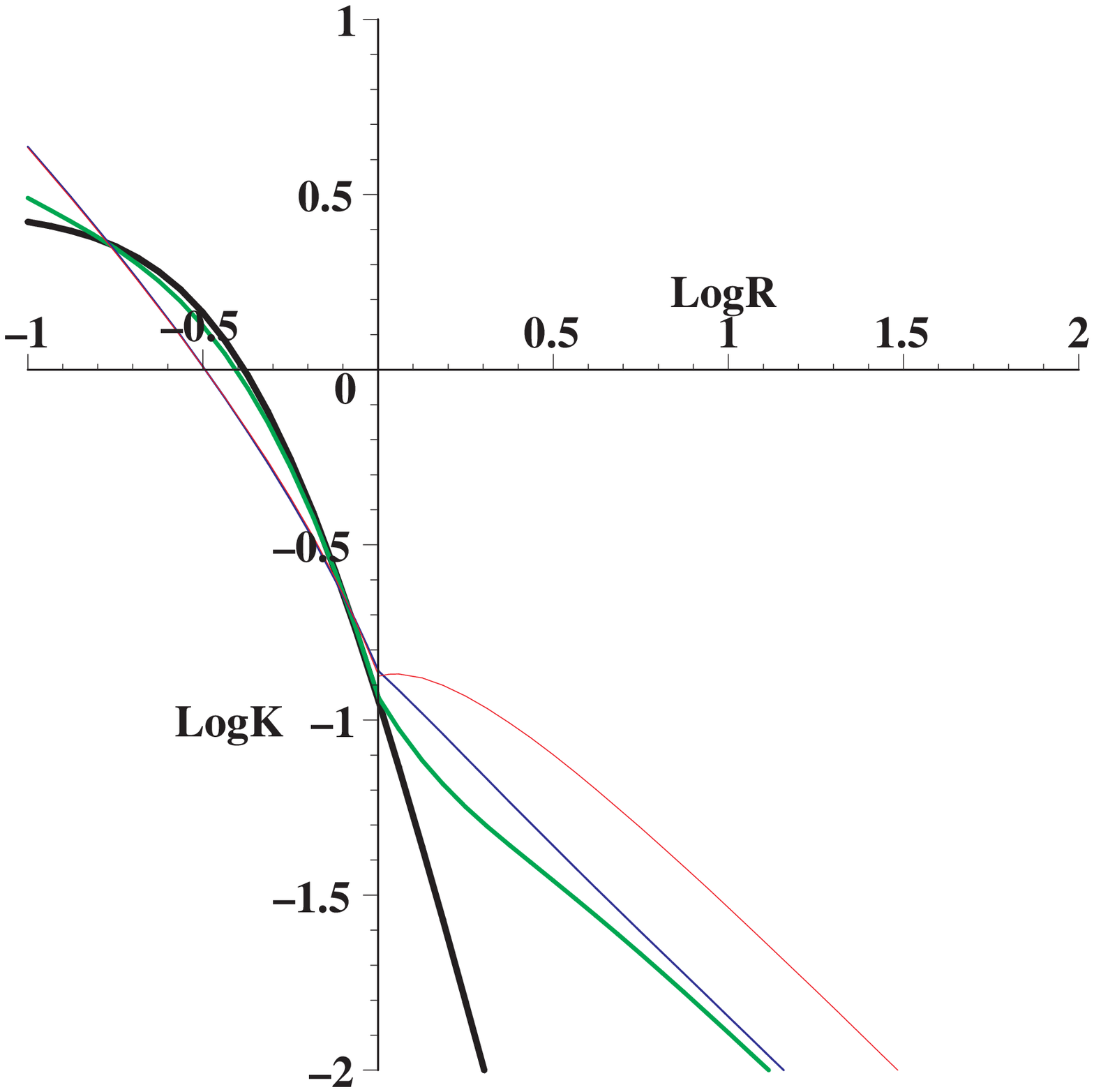}
\caption{Panel (a): The deflection strength of the lens $b=M(R)/R$ 
as a function of distance
from the lens; this is effectively a rotation curve of the lens.
From top right to down right, from thinner to thicker lines, the models are 
color coded as red (lensI), blue (lensII), green (lensIII) and black (lensIV).
Both Keplerian and flat rotation curve models are found.
Panel (b): Log-Log plot of the surface density profiles 
$\kappa(R)$ of the four lens models.
All models have very low $\kappa$ at the images near $-0.3 \le \log R \le 0$, 
consistent with a high $H_0$.
In all models except for lensIV (black), stars 
are dominated by dark matter halos 
just beyond the Einstein radius at $\log R \sim 0$. 
}
\label{vk}
\end{figure}

Among the acceptable models the rotation curve can be Keplerian or
flat (Fig.~\ref{vk}a), so lensing data plus $H_0$ cannot uniquely
specify the lens mass profile.
Among models in the literature, isothermal models and other simple
smooth models of dark matter halos of gravitational lenses often
predict a dimensionless time delay $H_0 \Delta t_{\rm obs}$ much too
small (e.g., Schechter et al. 1997) to be comfortable with the
observed time delays $\Delta t_{\rm obs}$ and the widely accepted
value of $H_0 \sim 70\kmsMpc$.  Naively speaking the high $H_0$
suggests a strangely small halo as compact as the stellar light
distribution (Kochanek 2002).  
But our analytical models suggest that there are
still many other options.  The high
$H_0$ implies that $\kappa$ is small $\sim 0.1-0.2$ at the images, but
this does not necessarily imply a rapidly falling density.
A high $H_0$ does not necessarily mean no dark halo, and models
with a flat rotation curve does not always yield a small $H_0$
(e.g., compare lensIII and lensIV in Figure~\ref{vk}b, both satisfy $h_0=0.7$).
We also comment that it will be difficult to determine the cosmology from
strong lensing data alone because the non-uniqueness in the lens models
implies that the combined parameter $\omega(H_0,\Omega,z_l,z_s)$ is
poorly constrained by the lensing data, even if $H_0$, $z_l$ and $z_s$ 
are given.

We thank the referee P. Saha for insightful 
comments on the cause of the degeneracy.
This work was supported by the National Science Foundation
of China under Grant No. 10003002 and a PPARC rolling
grant to Cambridge.  HSZ and BQ thank the Chinese Academy of Sciences
and the Royal Society respectively for a visiting fellowship, and the
host institutes for local hospitalities during their visits.

%\vfill\eject

\begin{deluxetable}{lcccccccccc} 
\tablecolumns{11} 
\tablewidth{0pc} 
\tablecaption{Plausible lens parameters to fit the images of 
a perfect Einstein cross and time delays with $H_0=100h_0=70\kmsMpc$
(cf. equations~(\ref{fitdelay}) and (\ref{fitimage})).} 
\tablehead{ 
\colhead{Model} 
& \colhead{Stars} & \colhead{Half-mass} & \colhead{Cusp} 
& \colhead{Halo} & \colhead{} & \colhead{Shear}
& \colhead{Conv.}  & \colhead{} 
& \colhead{Mass}  & \colhead{} 
\\
\colhead{} 
& \colhead{${M_0 \over R_E^2}$} & \colhead{$a={R_e \over R_E}$} & \colhead{$2-\alpha$} 
& \colhead{${b_0 \over R_E}$} & \colhead{$\delta$} & \colhead{$\gamma_1$}
& \colhead{$\kappa_*(R_E)$}  & \colhead{$\kappa_h(R_E)$} 
& \colhead{${M_*(<R_E)\over R_E^2}$}  & \colhead{${M_h(<R_E)\over R_E^2}$} 
}
\startdata
lensI &.2002& .5& 1.& .6220& .2& -.4445& .0222& .1110& .1334& .4220\\
lensII &.7110& 1.& 1.& .2988& .1& -.4458& .0888& .0494& .3555& .1988\\
lensIII &.4901& .5& 0.& .2759& .1& -.4317& .0784& .0379& .3921& .1759\\
lensIV &.7129& .5& 0.& .0013& 0.& -.4291& .1140& .0006& .5703& .0013\\
\enddata 
\end{deluxetable}

\end{document}